%% file: DES13S2cmm.tex
\documentclass[useAMS,usenatbib]{mn2e}

\usepackage{times}
\usepackage{graphics,epsf}
\usepackage{amsmath}                % American Mathematical Society package
\usepackage{amsfonts}               % American Mathematical Society fonts
\usepackage{amssymb}                % American Mathematical Society symbol
\usepackage{epsfig}                 % EPS figures
\usepackage{rotating}
\usepackage{color}
\usepackage{multirow}

\title[DES13S2cmm]{DES13S2cmm: The First Superluminous Supernova from the Dark Energy Survey}
\author[A. Papadopoulos et al.]{
\parbox[h]{\textwidth}{A. Papadopoulos$^{1}$\thanks{E-mail: andreas.papadopoulos@port.ac.uk}, C. B. D'Andrea$^{1}$, M. Sullivan$^{2}$, R. C. Nichol$^{1}$,
K. Barbary$^{3}$,
R. Biswas$^{4}$,
P. J. Brown$^{5}$,
R. A. Covarrubias$^{6,7}$
D.~A. Finley$^{8}$,
J.~A. Fischer$^{9}$,
R. J. Foley$^{7,10}$,
D. Goldstein$^{11,12}$,
R.~R. Gupta$^{4}$,
R. Kessler$^{13,14}$,
E. Kovacs$^{4}$,
S. E. Kuhlmann$^{4}$,
C. Lidman$^{15}$,
M. March$^{9}$,
P. E. Nugent$^{11,12}$,
M. Sako$^{9}$,
R.~C. Smith$^{16}$,
H. Spinka$^{4}$,
W. Wester$^{8}$, % SN WG ends here
T. M. C. Abbott$^{16}$,
F. Abdalla$^{17}$,
S.~S. Allam$^{6,18}$,
M. Banerji$^{17}$,
J.~P. Bernstein$^{4}$,
R.~A. Bernstein$^{19}$,
A. Carnero$^{20,21}$,
L.~N. da Costa$^{20,21}$,
D. L. DePoy$^{5}$,
S. Desai$^{22,23}$,
H.~T. Diehl$^{8}$,
T. Eifler$^{24}$,
A.~E. Evrard$^{25,26}$,
B. Flaugher$^{8}$,
J.~A. Frieman$^{8,13}$,
D. Gerdes$^{25}$,
D. Gruen$^{27,28}$,
K. Honscheid$^{29}$,
D. James$^{16}$,
K. Kuehn$^{15}$,
N. Kuropatkin$^{8}$,
O. Lahav$^{17}$,
M.~A.~G. Maia$^{20,21}$,
M. Makler$^{30}$,
J.~L. Marshall$^{5}$,
K.~W. Merritt$^{8}$,
C.~J. Miller$^{25,26}$,
R. Miquel$^{31,32}$,
R. Ogando$^{20,21}$,
A. A. Plazas$^{33}$
N.~A. Roe$^{12}$,
A. K. Romer$^{34}$,
E. Rykoff$^{35}$,
E. Sanchez$^{36}$,
B.~X. Santiago$^{21,37}$,
V. Scarpine$^{8}$,
M. Schubnell$^{25}$,
I. Sevilla$^{35}$,
M. Soares-Santos$^{8}$,
E. Suchyta$^{29}$,
M. Swanson$^{6}$,
G. Tarle$^{25}$,
J. Thaler$^{10}$,
D.~L.Tucker$^{8}$,
R. H. Wechsler$^{38}$,
J. Zuntz$^{39}$}}
\begin{document}

\maketitle

\label{firstpage}

\begin{abstract}
  We present DES13S2cmm, the first spectroscopically-confirmed
  superluminous supernova (SLSN) from the Dark Energy Survey (DES). We
  briefly discuss the data and search algorithm used to find this
  event in the first year of DES operations, and outline the
  spectroscopic data obtained from the European Southern Observatory
  (ESO) Very Large Telescope to confirm its redshift
  ($z=0.663\pm0.001$ based on the host-galaxy emission lines) and
  likely spectral type (type I). Using this redshift, we
  find $M^{peak}_U=-21.05^{+0.10}_{-0.09}$ for the peak, rest-frame
  $U$-band absolute magnitude, and find DES13S2cmm to be located in a
  faint, low metallicity (sub-solar), low stellar-mass host galaxy
  ($\log(\rmn{M}/\rmn{M}_{\odot})=9.3\pm0.3$); consistent with what is seen for 
  other SLSNe-I.  We compare the bolometric light curve of DES13S2cmm
  to fourteen similarly well-observed SLSNe-I in the literature and find it possesses one of the slowest declining tails (beyond $+30$ days rest frame past peak), and is the faintest at peak.
  Moreover, we find the bolometric light curves of all SLSNe-I studied herein possess a dispersion of only 0.2--0.3 magnitudes between $+25$ and $+30$ days after peak (rest frame) depending on redshift range studied; this could be important for `standardising' such supernovae, as is done with the more common type Ia. We fit the bolometric
  light curve of DES13S2cmm with two competing models for SLSNe-I --
  the radioactive decay of $^{56}$Ni, and a magnetar -- and find that
  while the magnetar is formally a better fit, neither model provides
  a compelling match to the data.  Although we are unable to
  conclusively differentiate between these two physical models for
  this particular SLSN-I, further DES observations of more SLSNe-I
  should break this degeneracy, especially if the light curves of
  SLSNe-I can be observed beyond 100 days in the rest frame of the
  supernova.
\end{abstract}

\begin{keywords}
surveys - stars: supernovae: general - stars: supernovae: DES13S2cmm 
\end{keywords}

\input{DES13S2cmm.sec1}

\input{DES13S2cmm.sec2}

\input{DES13S2cmm.sec3}

\input{DES13S2cmm.sec4}

\section{Conclusions}
\label{sec:conclusions}

In this paper we have presented DES13S2cmm, the first
spectroscopically-confirmed superluminous supernova (SLSN) from the
Dark Energy Survey. Using spectroscopic data obtained from the
ESO Very Large Telescope, we measured a redshift of $z=0.663\pm0.001$,
and assigned a classification of SLSN-I. However, we cannot exclude
the possibility that DES13S2cmm is a type R (Fig.~\ref{fig:spectra} and \ref{fig:Bol_lc})
if this is in fact a distinct SLSN type.

Using this redshift and correcting for Milky Way extinction, we find
the rest-frame $U$-band absolute magnitude of DES13S2cmm at peak to be
$M^{peak}_U=-21.05^{+0.10}_{-0.09}$, consistent with the SLSNe
definition of \citet{Gal-Yam12}.  Like other SLSNe \citep{lunnan14}, DES13S2cmm is
located in a faint low stellar-mass host galaxy
($\log(\rmn{M}/\rmn{M}_{\odot})=8.9\pm0.3$) with sub-solar metallicity.

In Fig.~\ref{fig:Bol_lc}, we compare the bolometric light curve of
DES13S12cmm to fourteen similarly well-observed SLSNe-I light curves in literature, and see that DES13S2cmm has one of the slowest declining tails (beyond $+30$ rest frame past peak) of all SLSNe-I studied herein, as well as likely being the faintest at peak. We further find that the dispersion between the bolometric light curves of all SLSNe-I shown in Fig.~\ref{fig:Bol_lc} has a minimum of $0.29$ magnitudes around $+30$ days past peak (rest frame). This reduces to $0.20$ magnitudes around $+25$ days if we remove the four SLSNe-I above $z\simeq1$ from the measurement because of the increased uncertainty in their rest-frame UV luminosities. This observation raises the tantalising possibility of
`standardising' these SLSNe like SNe Ia \citep{Quimby11,inserra14} and starting a new era of supernova cosmology \citep{king14}.  Further study is required to confirm if SLSNe can be standardised and investigate further the k-corrections of SLSNe, which would be needed to use SLSNe as robust distance indicators.

In Fig.~\ref{fig:Models}, we fit the bolometric light curve of
DES13S2cmm with two possible models for the power source of these
extreme events; the radioactive decay of $^{56}$Ni and a
magnetar \citep[e.g. see ][]{Inserra13}. 
We find that the $^{56}$Ni model is not a good fit to the bolometric
light curve, as the best-fitting model has a $\chi^2/$DoF$=2.7$ with an unrealistic fraction of $^{56}$Ni to ejected mass ($f_{\textrm{Ni}}=1$). The model does provide a 
relatively robust prediction for the overall $^{56}$Ni mass ($3.8-4.4 M_{\odot}$), as this is primarily
constrained by the peak luminosity of the light curve, but is unable to reproduce the late time ($t>30$ days) evolution of the light curve. The Magnetar model provides a smaller $\chi^2$ (for the same degrees of freedom) than $^{56}$Ni, but is likewise not a good fit to the data as
it is unable to fully reproduce the drop in luminosity within 30 days of explosion.

In the future, DES should find many more SLSN-I. Such data should allow us to further test
these two models, especially if we can measure the (rest frame)
light curves to $>100$ days (see Fig.~\ref{fig:Models} and Inserra et
al. 2013). Therefore, we have begun a new project entitled `SUrvey with Decam for Superluminous
Supernovae' (SUDSS), which will enable extended monitoring of some
DES SN fields (in addition to several non-DES fields) to over 8 months, greatly enhancing both
our ability to detect many more SLSNe and measure their long light curves. 

\section*{Acknowledgments}

The authors wish to thank Kate Maguire for her assistance with the ESO
VLT Directors Discretionary Time proposal. We also thank Cosimo Inserra and Stephen Smartt for helpful discussions regarding the classification, standardisation and k--corrections of superluminous supernovae. AP acknowledges the
financial support of SEPnet (www.sepnet.ac.uk) and the Faculty of
Technology of the University of Portsmouth. Likewise, CD and RN thank
the support of the Faculty of Technology of the University of
Portsmouth during this research, and MS acknowledges support from the
Royal Society and EU/FP7-ERC grant no [615929].  Part of TE's research was
carried out at JPL/Caltech, under a contract with NASA

Based on observations made with ESO telescopes at the La Silla Paranal Observatory under DDT program ID 292.D-5013

We are grateful for the extraordinary contributions of our CTIO
colleagues and the DES Camera, Commissioning and Science Verification
teams in achieving the excellent instrument and telescope conditions
that have made this work possible.  The success of this project also
relies critically on the expertise and dedication of the DES Data
Management organization. Funding for the DES Projects has been
provided by the U.S. Department of Energy, the U.S. National Science
Foundation, the Ministry of Science and Education of Spain, the
Science and Technology Facilities Council of the United Kingdom, the
Higher Education Funding Council for England, the National Center for
Supercomputing Applications at the University of Illinois at
Urbana-Champaign, the Kavli Institute of Cosmological Physics at the
University of Chicago, Financiadora de Estudos e Projetos, Funda{\c
  c}{\~a}o Carlos Chagas Filho de Amparo {\`a} Pesquisa do Estado do
Rio de Janeiro, Conselho Nacional de Desenvolvimento Cient{\'i}fico e
Tecnol{\'o}gico and the Minist{\'e}rio da Ci{\^e}ncia e Tecnologia,
the Deutsche Forschungsgemeinschaft and the Collaborating Institutions
in the Dark Energy Survey.

The Collaborating Institutions are Argonne National Laboratory, the
University of California at Santa Cruz, the University of Cambridge,
Centro de Investigaciones Energeticas, Medioambientales y
Tecnologicas-Madrid, the University of Chicago, University College
London, the DES-Brazil Consortium, the Eidgen{\"o}ssische Technische
Hochschule (ETH) Z{\"u}rich, Fermi National Accelerator Laboratory,
the University of Edinburgh, the University of Illinois at
Urbana-Champaign, the Institut de Ciencies de l'Espai (IEEC/CSIC), the
Institut de Fisica d'Altes Energies, Lawrence Berkeley National
Laboratory, the Ludwig-Maximilians Universit{\"a}t and the associated
Excellence Cluster Universe, the University of Michigan, the National
Optical Astronomy Observatory, the University of Nottingham, The Ohio
State University, the University of Pennsylvania, the University of
Portsmouth, SLAC National Accelerator Laboratory, Stanford University,
the University of Sussex, and Texas A\&M University.

This paper has gone through internal review by the DES collaboration. Please contact the author(s) to request access to research materials discussed in this paper.

\bibliographystyle{mn2e}
\bibliography{refs}

\section*{Affiliations} 

\small{
$^{1}$ Institute of Cosmology and Gravitation, Dennis Sciama Building, University of Portsmouth, Burnaby Road, Portsmouth, PO1 3FX, UK\\
$^{2}$ School of Physics and Astronomy, University of Southampton, Southampton, SO17 1BJ, UK\\
$^{3}$ Berkeley Center for Cosmological Physics, University of California at Berkeley, Berkeley, CA 94720, USA\\
$^{4}$ Argonne National Laboratory, 9700 S. Cass Avenue, Argonne, IL 60439, USA\\
$^{5}$ George P. \& Cynthia Woods Mitchell Institute for Fundamental Physics \& Astronomy,
Texas A. \& M. University, Dept. of Physics \& Astronomy, College Station, TX, USA\\
$^{6}$ National Center for Supercomputing Applications, 1205 West Clark St., Urbana, IL 61801, USA\\
$^{7}$Astronomy Department, University of Illinois at Urbana-Champaign, 1002 W.\ Green Street, Urbana, IL 61801, USA\\
$^{8}$ Fermi National Accelerator Laboratory, P.O. Box 500, Batavia, IL 60510, USA\\
$^{9}$Dept. of Physics \& Astronomy, University of Pennsylvania 209 South 33rd Street,Philadelphia, PA 19104, USA\\
$^{10}$ Department of Physics, University of Illinois Urbana-Champaign, 1110 W.\ Green Street, Urbana, IL 61801, USA \\
$^{11}$ Astronomy Dept., University of California at Berkeley, Berkeley, CA 94720, USA\\
$^{12}$ Lawrence Berkeley National Laboratory, 1 Cyclotron Road, Berkeley, CA 94720, USA\\
$^{13}$ Kavli Institute for Cosmological Physics, University of Chicago, Chicago, IL 60637, USA\\
$^{14}$ Department of Physics \& Astronomy, 5640 South Ellis Avenue, University of Chicago, Chicago, IL 60637, USA\\
$^{15}$ Australian Astronomical Observatory, PO Box 915, North Ryde NSW 1670, Australia\\
$^{16}$ Cerro Tololo Inter-American Observatory, National Optical Astronomy Observatory, Casilla 603, La Serena, Chile\\
$^{17}$ Department of Physics and Astronomy, University College London, London WC1E 6BT, UK\\
$^{18}$ Space Telescope Science Institute (STScI), 3700 San Martin Drive, Baltimore, MD 21218, USA\\
$^{19}$ Carnegie Observatories, 813 Santa Barbara St., Pasadena, CA 91101, USA\\
$^{20}$ Observat\'orio Nacional, Rua Gal. Jos\'e Cristino 77, Rio de Janeiro, RJ - 20921-400, Brazil\\
$^{21}$ Laborat\'orio Interinstitucional de e-Astronomia - LIneA, Rua Gal. Jos\'e Cristino 77, Rio de Janeiro, RJ - 20921-400, Brazil\\
$^{22}$ Department of Physics, Ludwig-Maximilians-Universit\"{a}t, Scheinerstr.\ 1, 81679 M\"{u}nchen, Germany\\
$^{23}$ Excellence Cluster Universe, Boltzmannstr.\ 2, 85748 Garching, Germany\\
$^{24}$ Jet Propulsion Laboratory, California Institute of Technology,4800 Oak Grove Drive, Pasadena, 91109 CA\\
$^{25}$ Department of Physics, University of Michigan, Ann Arbor, MI 48109, USA\\
$^{26}$ Department of Astronomy, University of Michigan, Ann Arbor, MI 48109, USA\\
$^{27}$ University Observatory Munich, Scheinerstrasse 1, 81679 Munich, Germany\\
$^{28}$ Max Planck Institute for Extraterrestrial Physics, Giessenbachstrasse, 85748 Garching, Germany\\
$^{29}$ Department of Physics, The Ohio State University, Columbus, OH 43210, USA\\
$^{30}$ ICRA, Centro Brasileiro de Pesquisas F\'isicas, Rua Dr. Xavier Sigaud 150, CEP 22290-180, Rio de Janeiro, RJ, Brazil\\
$^{31}$ Institut de F\'{\i}sica d'Altes Energies, Universitat Aut\`onoma de Barcelona, E-08193 Bellaterra, Barcelona, Spain\\
$^{32}$ Instituci\'o Catalana de Recerca i Estudis Avan\c{c}ats, E-08010 Barcelona, Spain\\
$^{33}$ Brookhaven National Laboratory, Dept of Physics, Bldg 510A, Upton, NY 11973\\
$^{34}$ Department of Physics and Astronomy, Pevensey Building, University of Sussex, Brighton, BN1 9QH, UK\\
$^{35}$ SLAC National Accelerator Laboratory, Menlo Park, CA 94025, USA\\
$^{36}$ Centro de Investigaciones Energ\'eticas, Medioambientales y Tecnol\'ogicas (CIEMAT), Avda. Complutense 40, Madrid, Spain\\
$^{37}$ Instituto de F\'\i sica, UFRGS, Caixa Postal 15051, Porto Alegre, RS - 91501-970, Brazil\\
$^{38}$ Kavli Institute. for Particle Astrophysics \& Cosmology, Stanford University, Stanford, CA 94305, USA\\
$^{39}$ Jodrell Bank Centre for Astrophysics, University of Manchester, Manchester M13 9PL, UK.
}

\bsp

\label{lastpage}

\end{document}

%% file: DES13S2cmm.sec1.tex
\section{Introduction}
\label{intro}

The last five years have seen the emergence of a new class of
ultra-bright stellar explosions: superluminous supernovae
\citep[SLSNe; for a review see][]{Gal-Yam12}, some 50 times brighter
than classical supernova (SN) types. Data on these rare and extreme
events are still sparse, with only $\sim$50 SLSN detections reported
in the literature. These events have generally been poorly studied
with incomplete imaging and spectroscopy, leaving many aspects of
their observational characteristics, and their physical nature,
unknown. 

Yet these SLSNe could play a key role in many diverse areas
of astrophysics: tracing the evolution of massive stars, driving feedback in
low mass galaxies at high redshift, and providing potential line-of-sight probes
of interstellar medium to their high-redshift hosts \citep{Berger12}.
SLSNe have also been detected to $z\sim4$ \citep{Cooke12}, far beyond
the reach of the current best cosmological probe, type Ia supernovae
(SNe Ia). If SLSNe can be standardised \citep[e.g.][]{inserra14}, as is done with SNe~Ia
\citep{tripp98,1993ApJ...413L.105P,Riess96}, then a new era of SN cosmology would be
possible, with the potential to accurately map the expansion rate of
the universe far into the epoch of deceleration \citep{delubac14}.

SLSNe have been divided into three possible types: SLSN-II, SLSN-I and
SLSN-R \citep[see Fig. 1 of][]{Gal-Yam12}. SLSNe-II show signs of interactions with CSM via narrow hydrogen lines
\citep[e.g.,][]{2007ApJ...659L..13O,2007ApJ...666.1116S}, and thus may simply represent
the bright end of a continuum of Type IIn SNe (although this is not well established).  

SLSN-I are
spectroscopically classified as hydrogen free \citep{Quimby11}, and
are possibly related to Type Ic SNe at late times
\citep{Pastorello10}, but normal methods of powering such SNe (e.g.,
the radioactive decay of $^{56}$Ni or energy from the gravitational
collapse of a massive star) do not appear to be able to simultaneously reproduce
their extreme brightness, slowly-rising light curves, and decay rates 
\citep{Inserra13}. Many alternative models have been proposed: the
injections of energy into a SN ejecta via the spin down of a young magnetar
\citep{Kasen10,Woosley10,Inserra13}; interaction of the SN ejecta with
a massive (3-5\,M$_{\odot}$) C/O-rich circumstellar material
\citep[CSM; e.g., ][]{2010arXiv1009.4353B}; or collisions between high-velocity shells 
generated from a pulsational pair-instability event \citep{Woosley07}. These explanations 
are still actively debated in the literature.
 
SLSNe-R are rare and characterised by possessing extremely long, slow-declining
light curves ($>200$\,days in the rest frame). SLSNe-R originally
appeared consistent with the death of $\gtrsim$100\,M$_{\odot}$ stars
via the pair instability mechanism \citep{Gal-Yam09}.  However, new
observations -- and the lack of significant spectral differences between 
SLSN-I and SLSN-R -- have challenged the notion that these classes are 
truly distinct \citep{Nicholl13} and maybe better described together as Type Ic SLSNe \citep{Inserra13}. 

In summary, the origin of the power source for SLSNe (of all types) remains unclear,
with the possibility that further sources of energy may be viable. A
more detailed understanding requires an increase in both the quantity
and quality of the data obtained on these events.  However, finding
more SLSNe (of any type) with existing transient searches is
challenging, especially in the local universe where the rates are only
$\simeq10^{-4}$ that of the core-collapse SN rate
\citep{Quimby13,2014arXiv1402.1631M} thus making it hard to assemble
large samples of events over reasonable timescales, even from large surveys. For example,
\citet{2014arXiv1402.1631M} detected only 10 SLSN candidates over
$0.3\le z\le1.4$ within the first year of the PanSTARRS1 Medium Deep
Survey.

Searches for SLSNe at higher redshifts may be more profitable as the rates
of SLSNe appear to rise by a factor of $\sim$10-15 at $z>1.5$
\citep{Cooke12}, perhaps tracking the increased cosmic star-formation
rate and/or decreasing cosmic metallicity: SLSNe are preferentially
found in faint, low-metallicity galaxies \citep{Neill11,Chen13,lunnan14}.
Moreover, the peak of a SLSN's energy output is located in the UV
region of the electromagnetic spectrum, which is redshifted to optical
wavelengths at high redshift.

In this paper, we outline our first search for SLSNe in the Dark
Energy Survey (DES). In Section~\ref{data}, we discuss details of DES
and our preliminary search, while in Section~\ref{sec:des13s2cmm}, we
describe the first SLSN detected in DES and provide details of its
properties, including our spectroscopic confirmation. In
Section~\ref{sec:results}, we compare our SLSN to other events in the
literature and explore the possible power sources for our event. We
conclude in Section~\ref{sec:conclusions}. Throughout this paper, we
assume a flat $\Lambda$-dominated cosmology with $\Omega_{M}=0.28$ and
$H_0=70$\,km\,s$^{-1}$\,Mpc$^{-1}$, consistent with recent
cosmological measurements \citep[e.g.][]{anderson14, betoule14}.

%% file: DES13S2cmm.sec2.tex
%SECTION 2 - DES 

\section{The Dark Energy Survey}
\label{data}

The Dark Energy Survey \citep[DES;][]{2005IJMPA..20.3121F} is a new
imaging survey of the southern sky focused on obtaining accurate
constraints on the equation-of-state of dark energy.  Observations are
carried out using the Dark Energy Camera
\citep[DECam;][]{2012SPIE.8446E..11F,2012PhPro..37.1332D} on the
4-metre Blanco Telescope at the Cerro Tololo Inter-American
Observatory (CTIO) in Chile.  DECam was successfully installed and
commissioned in 2012, and possesses a 3\,deg$^2$ field-of-view with 62
fully depleted, red-sensitive CCDs.

\subsection{DES Supernova Survey}
\label{sec:des-supernova-survey}

DES conducted science verification (SV) observations from November
2012 to February 2013, and then began full science operations in
mid-August 2013, with year one running until February 2014. 
DES is performing two surveys in parallel: a wide-field, multi-colour
($grizY$) survey of 5000\,deg$^2$ for the study of clusters of
galaxies, weak lensing and large scale structure; and a deeper,
cadenced multi-colour ($griz$) search for SNe. In this paper we focus
on this DES SN Survey, as outlined in \citet{Bernstein12}, which
surveys 30\,deg$^2$ over 10 DECam fields (two `deep', eight `shallow')
located in the XMM-LSS, ELAIS-S, CDFS and `Stripe82' regions.
As of November 2014, the DES SN Survey has already discovered over a thousand SN Ia candidates, with more than 70 spectroscopic classifications to date.

This rolling search for SNe Ia is also ideal for discovering other
transient phenomena including SLSNe. Using the light curve of
SNLS-06D4eu \citep[a SLSN-I; ][]{Howell13}, we calculate that DES
could detect such an event to $z=2.5$ given the limiting magnitudes
for the DES deep fields. For example, assuming a volumetric rate of
$32\,\rmn{Gpc}^{-3}\,\rmn{yr}^{-1}$ from \citet{Quimby13} and
correcting for time dilation, we estimate DES should find
approximately 3.5 SLSN-I events per DES season (five months) in the
deep fields and many more in the shallow fields (although at a lower
redshift).  This basic prediction is uncertain because: i) We have not
corrected for `edge effects' that will reduce the number of SLSNe with
significant light-curve coverage; ii) The volumetric rates themselves
may evolve with redshift \citep[e.g.,][]{Cooke12} and have large
uncertainties \citep{2014arXiv1402.1631M}; and iii) We have not
included any corrections for survey completeness.

\subsection{Selecting SLSNe candidates}
\label{sec:select-slsne-cand}

We began our search for SLSNe using the data products from the DES SN
Survey. All imaging data were de-trended and co-added using a standard
photometric reduction pipeline at the National Center for
Supercomputing Applications (NCSA) using the DES Data Management
system \citep[DESDM;][]{2012SPIE.8451E..0DM,2012ApJ...757...83D},
producing approximately 30 `search images' per field (in all filters)
over the duration of the five-month DES season. We perform difference
imaging on each of these search images, using deeper template images
for each field created from the co-addition of several epochs of data
obtained during the SV period in late 2012. Before differencing, the
search and template images were convolved to the same point spread
function (PSF).

Objects were selected from the difference images using
\textsc{SExtractor} \citep{1996A&AS..117..393B} v2.18.10, and previously
unknown transient candidates were identified and examined using
visual inspection (or `scanning') by DES team members. Over the course
of the first year of DES approximately $25,000$ new transient
candidates were selected in this manner, including many data artefacts
in addition to legitimate SN-like objects.  A machine learning algorithm (Goldstein et al., in preparation) 
was used to improve our efficiency of selecting real transients.  Transient candidates were 
photometrically classified using the Photometric SN IDentification
software \citep[PSNID;][]{Sako08,Sako11} to determine the likely SN
type based on fits to a series of SN templates.  These SN candidates
were stored in a database and prioritized for their spectroscopic
follow-up.

During the first season of DES, PSNID only included templates for the
normal SN types Ib/c, II and Ia, so a fully automated classification
of SLSN candidates was not possible. Therefore, for this first season
of DES, we searched for candidate SLSNe using the following criteria:
i) At least one month of multi-colour data, i.e., typically five to six
detections ($S/N>3.5$) in each of $griz$; ii) A low PSNID fit probability to any of the
standard SN sub-classes; iii) Located greater than one pixel from the
centroid of the host galaxy (to eliminate AGN); and iv) Peak observed brightness no fainter than one magnitude below that of its host galaxy.  These broad criteria will be
satisfied by most SLSNe, while also helping to eliminate many of the
possible contaminating sources.

Using this methodology, we selected 10 candidate SLSNe over the course
of the first season.  We secured a useable spectrum for
one of our SLSN candidates, DES13S2cmm, which is discussed in
detail in Section~\ref{sec:des13s2cmm}, while none of the other candidates
were spectroscopically confirmed. We are attempting to obtain spectroscopic redshifts from the host galaxies of these other SLSN candidates, though this will be challenging in the several cases where no host is detected in our template images.

%% file: DES13S2cmm.sec3.tex
\section{DES13S2cmm}
\label{sec:des13s2cmm}

\begin{figure*}
	\includegraphics[width=0.99\textwidth,clip]{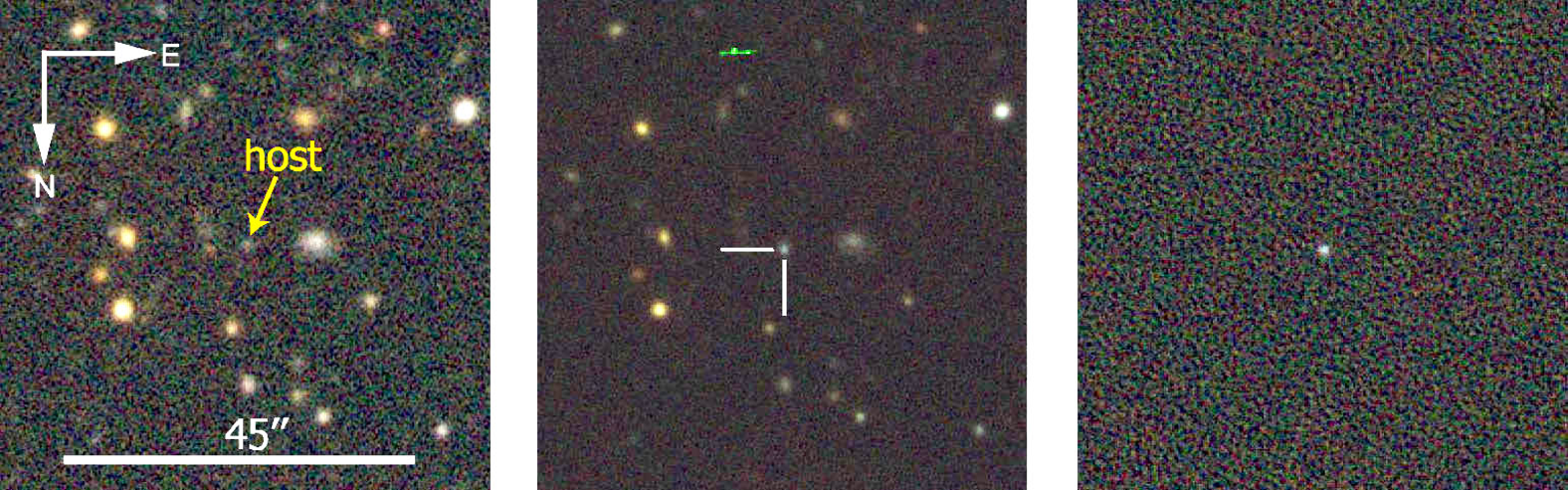}
	\caption{Colour images (\textit{riz}) of the field surrounding
          DES13S2cmm. \textit{Left:} Deep template image created from
          the co-addition of several epochs of data obtained during
          the DES Science Verification period in late 2012; the likely
          host galaxy is indicated by the arrow. \textit{Centre:} The
          search image taken close to maximum light (28-Sept-2013)
          with the position of DES13S2cmm indicated. \textit{Right:}
          The difference of the two previous images, clearly
          identifying DES13S2cmm at center.  This image demonstrates
          the quality of difference images used for discovery and
          monitoring during the DES observing season.}
\label{fig:triplet}
\end{figure*}

In Fig. \ref{fig:triplet} we show DES \textit{riz} imaging data at the
position of DES13S2cmm: pre-explosion, post-explosion, and their
difference image.  DES13S2cmm is located at
$\rmn{RA}(2000)=02^{\rmn{h}} 42^{\rmn{m}} 32\fs82$,
$\rmn{Dec.}~(2000)=-01\degr 21\arcmin 30\farcs 1$, and was first
detected on 2013 August 27 in the DES SN `S2' field (a `shallow'
field) located in the Stripe 82 region. DES transient names are formatted
following the convention \hbox{DESYYFFaaaa}, where YY are the last two digits of the year in
which the observing season began (13), FF is the two character field name (S2), and the final
characters (all letters, maximum of 4) provide a running candidate
identification, unique within an observing season, as is traditional
in SN astronomy (cmm).

%--------------------------------------------------------------------------------------------------------------------------------------------------------

\subsection{Light Curve}
\label{sec:light-curve}

\begin{figure*}
	\includegraphics[width=0.9\textwidth,clip]{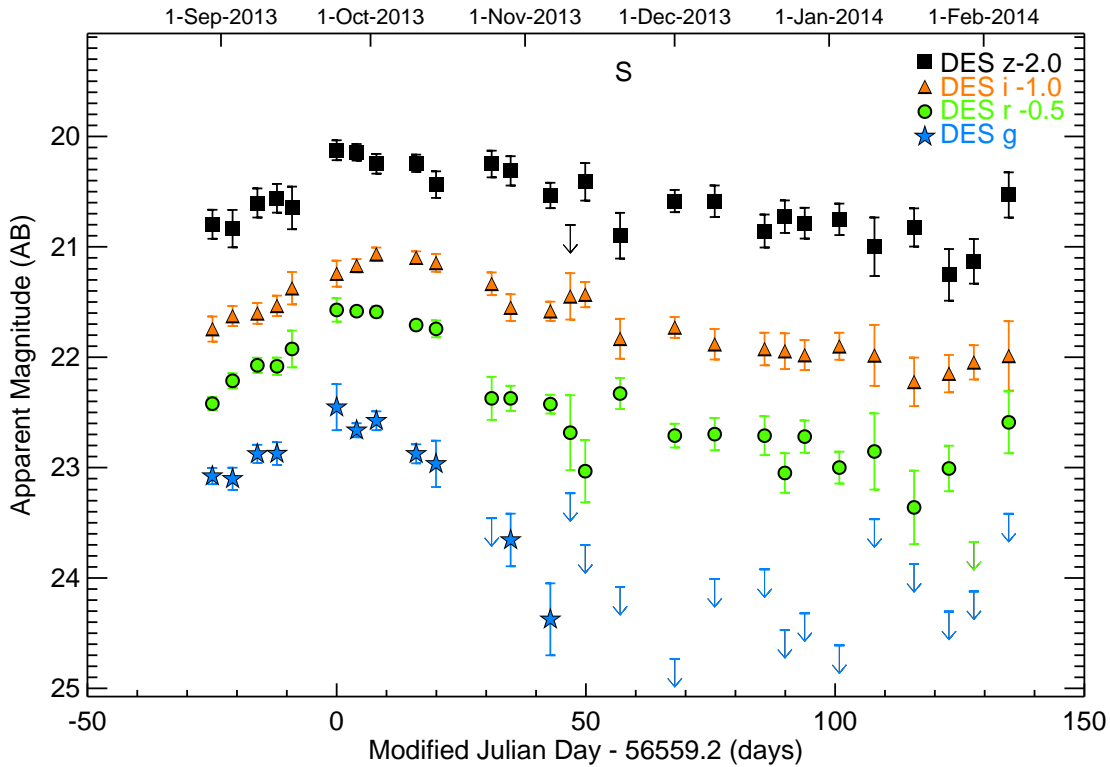}
	\caption{The observed AB magnitude light curve of DES13S2cmm
          in the four DES SN search filters ($griz$) as a function of
          the observed phase. `S' denotes the epoch when the ESO VLT
          spectrum was obtained. The error-bars represent 1-$\sigma$
          uncertainties, and the arrow symbols are 3-$\sigma$ upper
          limits. The phase relative to peak was defined using the
          observed-frame $r$-band. We have artificially offset (in
          magnitudes) the $r$, $i$ and $z$ band data for clarity.}
\label{fig:light_curve}
\end{figure*}

In Fig.~\ref{fig:light_curve} we present the multi-colour light curve
for DES13S2cmm constructed from the first year of DES observations. We
have not included upper limits from the earlier, pre-explosion epochs
available from the DES SV period in 2012-2013, as these data were
taken hundreds of days prior to the data shown in
Fig.~\ref{fig:light_curve} and are already used to create the
reference template image for this field used in the difference imaging
detection of the SNe. 

We use the photometric pipeline from
\citet{Sullivan11} \citep[also used in][]{Maguire12,Ofek13} to reduce
the DES photometric data for DES13S2cmm. This pipeline is based on
difference imaging, subtracting a deep, good seeing, pre-explosion
reference image from each frame the SN is present in. The photometry
is then measured from the differenced image using a PSF-fitting method, with the PSF determined
from nearby field stars in the unsubtracted image. This average PSF is
then fit at the position of the SN event, weighting each pixel
according to Poisson statistics, yielding a flux, and flux error, for
the SN. 

We calibrate the flux measurements of DES13S2cmm to a set of tertiary
standard stars produced by the DES collaboration (Wyatt et al., in preparation), intended to be in
the AB photometric system \citep{Oke83}.
The photometry shown in Fig.~\ref{fig:light_curve} has also been
checked against the standard DES detection photometry, used to find
and classify all candidates for follow-up, and found to be in good
agreement. The light-curve data are presented in
Table~\ref{tab:photometry_table}.

We correct for Galactic extinction using the maps of \citet{SF11}, which estimate ${\rm E(B-V)}=0.028$ at the location of DES13S2cmm, 
leading to extinction values of 0.123, 0.070, 0.051 and 0.039 magnitudes in the DES $griz$ filters respectively.  We do not include these
corrections in the values reported in Table~\ref{tab:photometry_table}, but do correct for extinction in all figures and analysis herein, 
including Figure~\ref{fig:light_curve}.

\begin{table*}
\begin{center}
  \caption{Light curve data for DES13S2cmm used in
    Fig.~\ref{fig:light_curve}. We provide the observed calendar date,
    modified Julian Day (MJD), observed phase relative to peak MJD
    (defined relative to peak flux in the \textit{r}-band), and fluxes (not corrected for Galactic extinction)
    with 1-$\sigma$ uncertainties in the DES \textit{griz} passbands. These fluxes
    can be converted to AB magnitudes using
    $\rmn{mag}_{\rmn{AB}}=-2.5\log_{10}\left(f\right)+31$. }
    \begin{tabular}{ c | c | c || r @{$\pm$}l |  r @{$\pm$} l  |  r @{$\pm$}  l |  r @{$\pm$} l |}
		\hline \hline
\multicolumn{1}{|c|}{Date} &\multicolumn{1}{|c|}{MJD} & \multicolumn{1}{|c|}{Phase} & \multicolumn{2}{|c|}{$f_g$} & \multicolumn{2}{|c|}{$f_r$} & \multicolumn{2}{|c|}{$f_i$} & \multicolumn{2}{|c|}{$f_z$} \\ 
&&(days)&\multicolumn{2}{|c|}{}&\multicolumn{2}{|c|}{}&\multicolumn{2}{|c|}{}&\multicolumn{2}{|c|}{}\\
\hline		
% Date               &   MJD	    &  Phase   &   g       &           &  r          &           &  i           &           &   z        &          \\ 	
30-AUG-2013    &  56534.3   &  -25   &    1138 &  77 &    1705 &  94 &    1730 & 181 &    1913 & 231  \\
 3-SEP-2013    &  56538.3   &  -21   &    1112 & 103 &    2063 & 129 &    2234 & 186 &    1845 & 288  \\
 8-SEP-2013    &  56543.3   &  -16   &    1369 & 104 &    2348 & 148 &    2285 & 198 &    2285 & 277  \\
12-SEP-2013    &  56547.2   &  -12   &    1373 & 130 &    2329 & 167 &    2433 & 207 &    2376 & 284  \\
15-SEP-2013    &  56550.2   &   -9   &              &       &       2691 & 413 &    2818 & 380 &    2193 & 389  \\
24-SEP-2013    &  56559.2   &    0   &    2022 & 388 &    3727 & 363 &    3183 & 346 &    3548 & 295  \\
28-SEP-2013    &  56563.2   &    4   &    1668 &  97 &    3688 & 133 &    3400 & 187 &    3488 & 241  \\
 2-OCT-2013    &  56567.2   &    8   &    1805 & 141 &    3666 & 166 &    3742 & 209 &    3169 & 261  \\
10-OCT-2013    &  56575.2   &   16   &    1370 & 108 &    3285 & 137 &    3636 & 192 &    3182 & 230  \\
14-OCT-2013    &  56579.1   &   20   &    1260 & 243 &    3182 & 223 &    3482 & 259 &    2663 & 295  \\
25-OCT-2013    &  56590.3   &   31   &     886 & 346 &    1782 & 319 &    2924 & 277 &    3163 & 352  \\
29-OCT-2013    &  56594.1   &   35   &     667 & 146 &    1782 & 187 &    2398 & 267 &    2987 & 366  \\
 6-NOV-2013    &  56602.1   &   43   &     344 & 103 &    1698 & 132 &    2326 & 185 &    2433 & 257  \\
10-NOV-2013    &  56606.1   &   47   &     803 & 426 &    1338 & 418 &    2633 & 510 &    1817 & 634  \\
13-NOV-2013    &  56609.1   &   50   &    -312 & 276 &     970 & 252 &    2672 & 281 &    2726 & 426  \\
20-NOV-2013    &  56616.1   &   57   &     347 & 195 &    1855 & 239 &    1848 & 307 &    1739 & 331  \\
 1-DEC-2013    &  56627.0   &   68   &     184 & 107 &    1306 & 129 &    2032 & 176 &    2322 & 216  \\
 9-DEC-2013    &  56635.0   &   76   &     259 & 208 &    1322 & 178 &    1766 & 225 &    2318 & 304  \\
19-DEC-2013    &  56645.1   &   86   &     302 & 226 &    1305 & 211 &    1697 & 230 &    1808 & 248  \\
23-DEC-2013    &  56649.1   &   90   &     261 & 136 &     956 & 157 &    1667 & 248 &    2038 & 276  \\
27-DEC-2013    &  56653.1   &   94   &      41 & 156 &    1295 & 174 &    1613 & 201 &    1930 & 248  \\
 3-JAN-2014    &  56660.0   &  101   &     252 & 119 &    1000 & 132 &    1736 & 196 &    1993 & 261  \\
10-JAN-2014    &  56667.1   &  108   &     310 & 343 &    1144 & 364 &    1609 & 408 &    1586 & 387  \\
18-JAN-2014    &  56675.0   &  116   &     -75 & 236 &     717 & 220 &    1290 & 261 &    1863 & 297  \\
25-JAN-2014    &  56682.0   &  123   &     306 & 158 &     992 & 188 &    1382 & 215 &    1254 & 270  \\
30-JAN-2014    &  56687.0   &  128   &     345 & 188 &     523 & 179 &    1518 & 217 &    1404 & 261  \\
 6-FEB-2014    &  56694.0   &  135   &    -346 & 359 &    1459 & 376 &    1602 & 465 &    2443 & 461  \\
\hline
     \end{tabular}
  \label{tab:photometry_table}
\end{center}
\end{table*}

%--------------------------------------------------------------------------------------------------------------------------------------------------------

\subsection{Spectroscopy}
\label{sec:spectrum}

On 2013 November 12, we requested Director's Discretionary Time
at the European Southern Observatory (ESO) Very Large Telescope (VLT)
to observe DES13S2cmm, and were awarded target-of-opportunity time within days.
The closeness of the object to the moon and instrument scheduling
issues conspired to delay observations until 2013 November 21, at
which point DES13S2cmm was approximately 30 days after peak brightness
(rest frame). A spectrum was obtained with the FOcal Reducer and low
dispersion Spectrograph \citep[FORS2][]{Appenzeller98} using the
GRIS\_300I+11 grism, the OG590 order blocker, and a 1\arcsec\ slit,
with an exposure time of 3600s (3$\times$1200s). This configuration
provided an effective wavelength coverage of $5950$\,\AA\ to
$9400$\,\AA.

The reduction of the FORS2 data followed standard procedures using the
Image Reduction and Analysis Facility\footnote{The Image Reduction and
  Analysis Facility (\textsc{iraf}) is distributed by the National
  Optical Astronomy Observatories, which are operated by the
  Association of Universities for Research in Astronomy, Inc., under
  cooperative agreement with the National Science Foundation.}
environment (v2.16), using the pipeline described in \citet{Ellis08}. This
pipeline includes an optimal two-dimensional (2D) sky subtraction
technique as outlined in \citet{Kelson03}, subtracting a 2D sky frame
constructed from a sub-pixel sampling of the background spectrum and a
knowledge of the wavelength distortions determined from 2D arc
comparison frames. The extracted 1D spectra were then scaled to the
same flux level, the host-galaxy emission lines interpolated over, and the
spectra combined using a weighted mean and the uncertainties in
the extracted spectra.

\begin{figure*}
	\includegraphics[width=0.95\textwidth,clip]{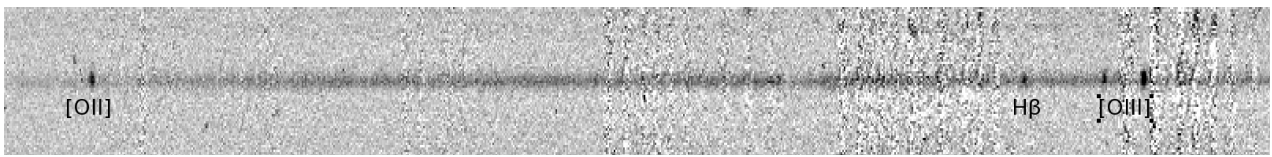}
	\caption{The reduced VLT FORS2 two-dimensional spectrum for DES13S2cmm. The key nebular emission lines used to derive the redshift of the host galaxy, and therefore supernova, are labelled.}
	\label{2Dspectrum}
\end{figure*}

In Fig \ref{2Dspectrum} we present our reduced 2D FORS2 spectrum, where wavelength increases to the right along the horizontal axis and the vertical axis is distance along the slit.  We highlight the obvious narrow nebular emission lines in the host galaxy of DES13S2cmm, specifically [O\,\textsc{ii}]~$\lambda3727$\,\AA, H$\beta$ $\lambda4861$\,\AA, and [O\,\textsc{iii}]~$\lambda4959,5007$\,\AA.  We assume these lines originate from the underlying host galaxy as their spatial extent along the vertical axis is greater than the observed trace of the main supernova spectrum. This observation does not exclude the possibility that such emission lines come from the supernova, but it is difficult to address this issue with our low resolution, and low signal-to-noise, data. We note the observed emission lines are not significantly broadened or asymmetrical, as witnessed in type IIn supernovae. 

We see no evidence of any further emission lines, from the SN and/or host galaxy. Therefore, it is unlikely DES13S2cmm is a type II supernova (normal or superluminous). 

Using these nebular emission lines, we determine a redshift of $z=0.663\pm0.001$ for the host galaxy of
DES13S2cmm, where the uncertainty is derived from the
wavelength dispersion between individual emission lines.  
This measurement is consistent with the redshift of
the SN based on identifications of the broad absorption features seen
in the spectrum.  As the redshift derived from the host-galaxy spectrum is significantly 
more precise, henceforth we adopt this value as the redshift to DES13S2cmm.

The spectral classification for DES13S2cmm is based on the
\textsc{superfit} program \citep{Howell05}, which compares 
observed spectra of SNe to a library of template SN spectra (via
$\chi^2$ minimisation) while accounting for the possibility of host-galaxy
contamination in the data. We added as template spectra of SLSNe-I
(PTF09atu, PTF09cnd, PTF10nmn) and SLSN-R (SN\,2007bi) to
\textsc{superfit} to facilitate our classification.  
We found that the highest-ranked fits to
DES13S2cmm were these SLSNe templates, preferred over templates of normal SN types (e.g., SN Ia). The SLSNe-I templates were slightly better fits to DES13S2cmm than SN\,2007bi (SLSN-R), although this may be due to not having a template for SN\,2007bi at a similar phase to our spectrum for DES13S2cmm.

\begin{figure*}
	\includegraphics[width=0.9\textwidth,clip]{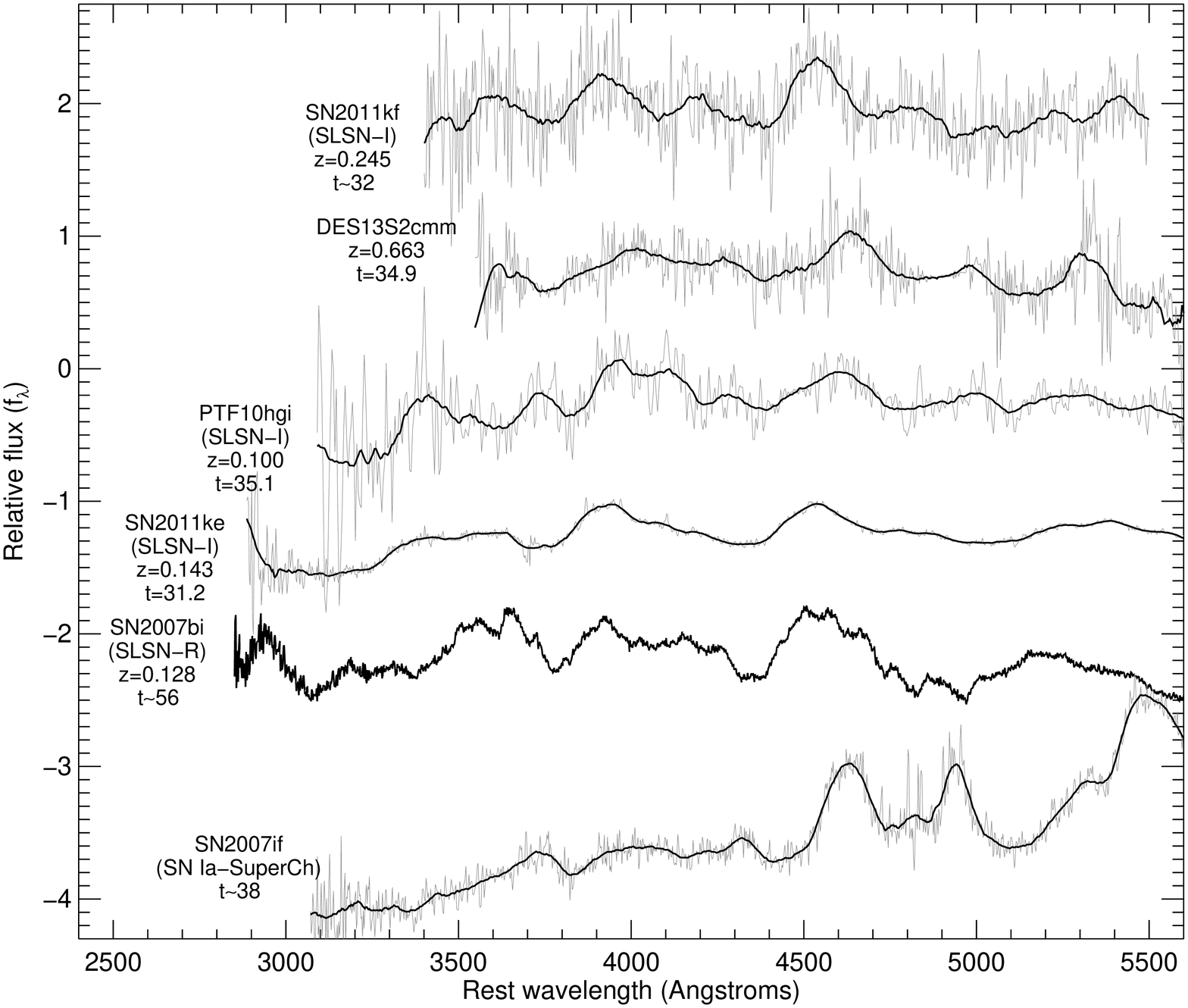}
	\caption{The VLT FORS2 spectrum of DES13S2cmm (second from
          top) compared to other SLSNe of similar phase in their light
          curves. From top to bottom, we also show SN\,2011kf
          \citep{Inserra13}, PTF10hgi \citep{Quimby11}, SN\,2011ke
          \citep{Inserra13} and SN\,2007bi\citep[classified as Type R in][]{Gal-Yam09}; as discussed in 
          Section \ref{sec:spectrum}, \textsc{superfit} identifies these SLSNe as 
          some of the best-fitting spectral templates to DES13S2cmm.  In each
          case, the light grey shows the raw data and the solid line
          is a Savitsky-Golay smoothed version of the spectrum. The
          spectra are labelled with the SN name, SLSN type (except
          DES13S2cmm), redshift, and the phase when the spectrum
          was taken. We show the spectrum of the `super-Chandrasekhar'
          SN Ia SN\,2007if as discussed in \citet{scalzo10}. Note that the
          spectrum for 2007bi, while at a later phase than the other data, is the 
          earliest spectral observation of this SN.}
	\label{fig:spectra}
\end{figure*}

We show our spectrum of DES13S2cmm in Fig.~\ref{fig:spectra}, alongside
spectra from other well-studied SLSNe in the literature at a similar
phase in their light curves \citep{Quimby11,Inserra13,Gal-Yam09}.
There is a deficit of spectra for SLSNe in the literature at these
late phases for a robust comparison.  However, the broad resemblance
between the spectra of PTF10hgi, SN\,2011kf, SN\,2011ke and
DES13S2cmm, all taken at approximately $+30$ days past peak, suggests
DES13S2cmm is a similar object to these other SLSNe-I. The spectrum of
DES13S2cmm is also similar to the spectrum of SN\,2007bi (a SLSN-R in \citealp{Gal-Yam09}), although the differences between Type I and Type R SLSNe remains unclear, and either classification would be interesting given the sparseness of the
data. We also show in Fig.~\ref{fig:spectra} the spectrum of
SN\,2007if, a `super-Chandrasekhar' SN Ia \citep{scalzo10}, an
over-luminous ($M_V = -20.4$) and likely thermonuclear SN located in a
faint host galaxy.  This comparison clearly shows that DES13S2cmm is
unlikely to be a super-Chandrasekhar SN Ia.

%--------------------------------------------------------------------------------------------------------------------------------------------------------

\subsection{Host-galaxy properties}
\label{sec:host-galaxy}

The host galaxy of DES13S2cmm is detected in our reference images that
contain no SN light. Using \textsc{sextractor} in dual-image mode and the $i$-band image as the detection image, we
measure host-galaxy magnitudes of $g=24.24\pm0.13$, 
$r=23.65\pm0.10$, $i=23.31\pm0.10$ and $z=23.26\pm0.17$ (AB \textsc{mag\_auto}
magnitudes).  The large photometric errors are due to the relatively
shallow depth, and poor seeing, of the template image from the DES SV data -- we cannot use the DES first-year data for
the host photometry as all epochs of this data contain some light from
DES13S2cmm. 

The estimated photometric redshift of the host galaxy at the time of discovery was 
$z_{\rmn{photo}}=0.86\pm0.13$ (68 percentile error).  This photometric
redshift was derived by the DESDM neural network photo-z module code, with uncertainties estimated 
from the nearest neighbour method, using data taken during SV \citep{sanchez14}.  
Photometric redshifts (when available) are
used when selecting SLSN candidates for spectroscopic follow-up, and
for DES13S2cmm this redshift implied the event was brighter than
$M^{\rmn{peak}}_U =-21$ \citep[typical for SLSNe; ][]{Gal-Yam12}.  
Since the start of the first DES observing season the photometric catalog derived from SV data has improved, and presently the DESDM photo-z for the host of DES13S2cmm has improved to $z_{\rmn{photo}}=0.71\pm0.06$, which is in better agreement with our measured spectroscopic redshift ($z=0.663\pm0.001$).

Using the DES host-galaxy photometry and the spectroscopic redshift from VLT, we estimate that the stellar mass of the host galaxy is
$\log(M/M_{\odot})=9.3\pm0.3$ using stellar population models from
PEGASE.2 \citep{1997A&A...326..950F}, and
$\log(M/M_{\odot})=9.0\pm0.3$ using the stellar population templates
from \citet{2005MNRAS.362..799M} (both corrected for Milky Way extinction and fixing the redshift of the templates to $z=0.663$). Such a low stellar-mass host galaxy
is consistent with the findings for other SLSNe \citep{Neill11,Chen13,lunnan14}.

We estimate the host-galaxy metallicity from our VLT spectrum using the double-valued metallicity indicator $R_{23}$, defined as $([\textrm{O}\textsc{II}]~\lambda 3727 + [\textrm{O}\textsc{III}]~\lambda\lambda 4959, 5007)/H\beta$.  Fluxes and uncertainties are derived individually for each emission line from fits of a gaussian plus a first-order polynomial, with the assumption that there is no contamination of the galaxy emission lines from the SN.  We compute metallicities using the calibrations of \citet{KK04} and \citet{M91}, and use the formulae derived in \citet{KE08} to convert these metallicities into the calibration of \citet{KD02} -- a step that allows for direct comparison between different methods.  We note that since $H\alpha$ and $\textrm{N}\textsc{II}~\lambda6584$ are redshifted to wavelengths beyond our spectral coverage, we cannot determine the branch of the $R_{23}$ function that the metallicity lies on.  However, the derived value for $R_{23}$ is close to its theoretical maximum, which minimizes the difference in metallicity estimates between the two branches.  

Assuming the lower (upper) branch, we find a metallicity ($12+\log$[O/H]) of 8.30 (8.38) from \citet{M91} and 8.30 (8.42) from \citet{KK04}, which is consistent with the median metallicity of $8.35$ found by \citet{lunnan14} for a sample of 31 SLSNe host galaxies.  The total uncertainty is dominated by the calibration ($\sim0.15$ dex), while the statistical uncertainty and the scatter induced by the conversion to \citet{KD02} are comparatively negligible.  We thus find a host-galaxy abundance ratio that is distinctly sub-solar \citep[8.69;][]{Asplund09}, though not as extremely low as has been seen for other SLSNe, such as SN2010gx ($12+\log$[O/H]=7.5;~\citealp{Chen13}).

%% file: DES13S2cmm.sec4.tex
\section{Results}
\label{sec:results}

\subsection{Bolometric light-curve of DES13S2cmm}
\label{peakM}

\begin{figure*}
	\includegraphics[width=0.9\textwidth,clip]{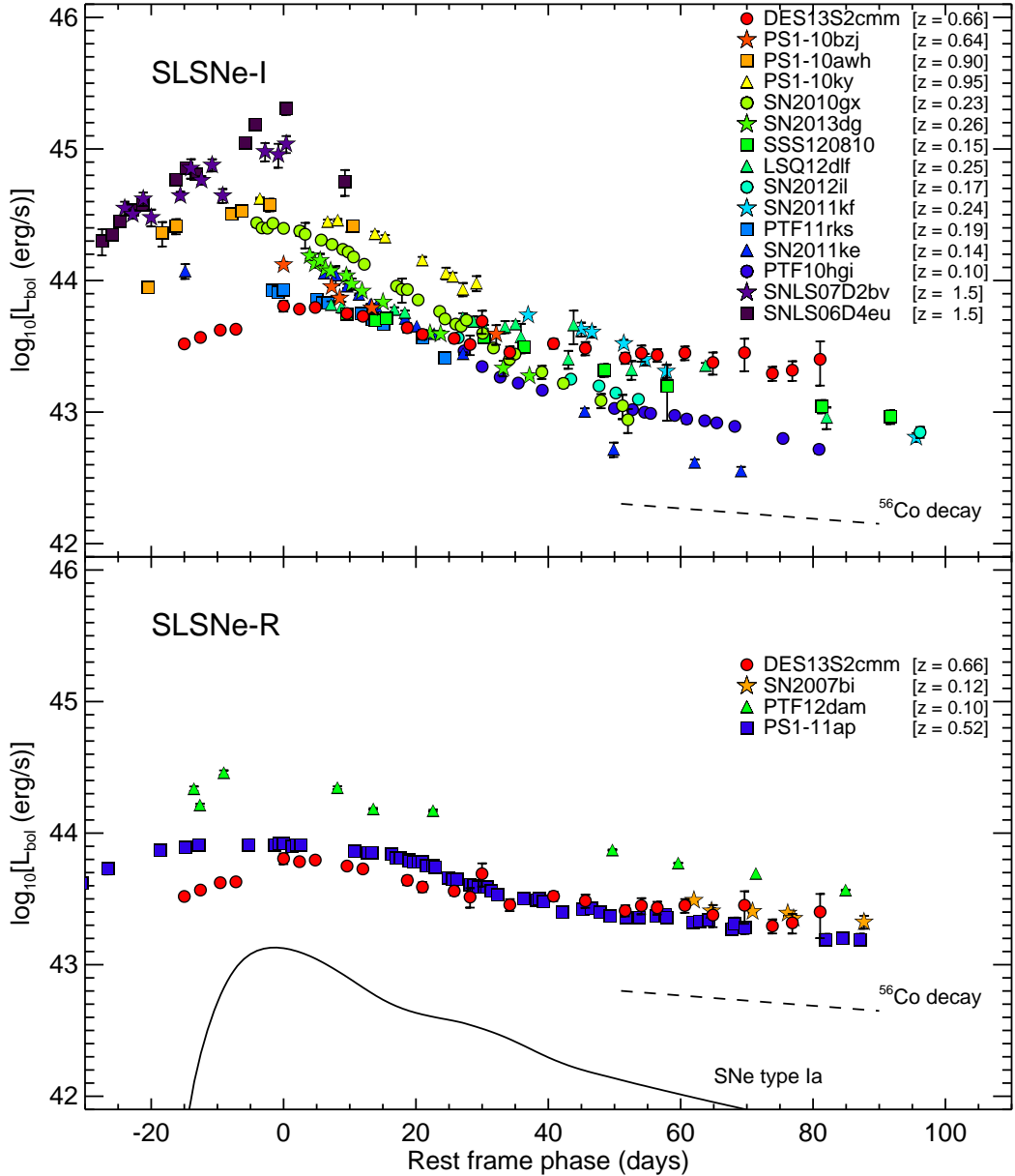}
	\caption{The bolometric light curve of DES13S2cmm (red
          circles) compared to other SLSNe in the literature (see text for explanation). These bolometric
          light curves were constructed by fitting a single black-body
          spectrum to the multicolour data available per epoch
          (requiring at least three photometric measurements per
          epoch), and then by integrating the black-body spectral
          energy distribution to produce the bolometric luminosity per
          rest-frame epoch (described in detail in
          Section~\ref{Bol_lc}). We show statistical errors on the blackbody fits to each epoch of these light curves, but in most cases this error is 
          smaller than the plotting symbol. We stress that these errors do not account for systematic uncertainties 
          involved in the bolometric calculation (e.g., assuming a blackbody
          for the spectral energy distribution of SLSNe, especially for the UV part of the rest-frame spectral energy distribution).  The phase relative to peak (y-axis) for the literature SLSNe is taken directly from the
          published individual studies and not recalculated herein. The dashed lines shown in both
          panels represent the expected $^{56}$Co decay rate at late times. In the lower panel, we show the bolometric light curve expected for a typical Type Ia SNe (solid line) to again illustrate the extraordinary brightness, and extended duration, of these SLSNe}
	\label{fig:Bol_lc}
\end{figure*}

We show in Fig.~\ref{fig:Bol_lc} the bolometric light curve of
DES13S2cmm. This single light curve was constructed by fitting a
single black-body spectrum to the DES multi-colour data (requiring
measurements in a minimum of three photometric bandpasses for a fit),
on each epoch in the light curve (Table 1). In detail, we redshift
black-body spectra to the observer frame and integrate these spectra
through the response functions of the DES filters. We then determine
the best-fitting parameters for a black body to the
extinction-corrected photometry and integrate this spectrum in the
rest frame to obtain the bolometric luminosity per epoch.  This
procedure benefits from the homogeneity of the DES data as we possess
approximately equally-spaced epochs across the whole light curve --
typically in four DES passbands -- due to our rolling search. The
errors on the bolometric luminosities were calculated from the fitting
uncertainties on the black-body spectra.

We visually checked our best-fit blackbody curves for each epoch of the DES13S2cmm light curve against the observed $griz$ photometry and found them to be reasonable within the errors. The largest discrepancies were seen in the g-band where some epochs were below the fitted curves, possibly due to UV absorption in the underlying spectral energy distributions relative to a blackbody curve. Inserra et al. (2013) compared the predicted UV flux (from a blackbody curve) for two SLSNe with SWIFT UV observations, and found the fluxes were consistent within the error. Therefore we make no additional correction 
to the derived bolometric light curve, but note that quantifiable systematic errors
in the bolometric luminosity require spectral templates, which in turn requires 
more spectral data of SLSNe than is currently publicly available.

Using the bolometric light curve of DES13S2cmm, we can determine the
peak luminosity of the event and compare it to the definition of a
SLSNe in \citet{Gal-Yam12}. This was achieved by integrating the
best-fitting black-body spectra through a standard $U$-band filter
response function (on the Vega system) to obtain the rest-frame absolute magnitude in
$U$-band ($M_U$). We find the 2013 September 24 epoch ($t=0$ relative
phase in Table 1) gives the largest (peak) absolute $U$-band magnitude
of $M^{peak}_U=-21.05^{+0.10}_{-0.09}$, which is consistent with the
SLSNe threshold of $M^{peak}_U<-21$ in \citet{Gal-Yam12}. 

We check this result using a model-independent estimate based on the observed $r$-band peak magnitude.
At the redshift of DES13S2cmm, the observer-frame $r$ band maps into the rest frame as a synthetic filter
with an effective wavelength of $\approx3800$\AA, slightly redward of the standard $U$-band.  Defining a synthetic filter 
as such means the k-correction is independent of the SN spectral energy distribution, and since we are in the AB system
the correction is simply $2.5\log(1+z)$.  We apply this k-correction to the observed $r$-band peak magnitude, as well as 
corrections for Galactic extinction and distance modulus, to yield an estimated peak absolute magnitude
of $-20.47$ (AB), or $-20.74$ (Vega) in our synthetic filter (where we have computed the AB to Vega conversion for our filter).  
While this is fainter than our blackbody-based estimate, Vega magnitudes are sensitive to the exact 
shape of the filter in this region due to the jump in Vega from 3700-3900\AA; thus even the small difference in effective 
wavelength between $U$ band and our synthetic filter can result in significant offsets.  The absolute magnitude in our 
synthetic filter from the best-fitting blackbody method yields $M_{\rm synthetic}=-20.59$, proving that our 
model-independent estimate is consistent with our results from the blackbody fitting.

We note that the peak magnitude (Vega) for DES13S2cmm classifies it
as a SLSN, provided the magnitude is intrinsic to the SN and not
enhanced by other effects.  Visual inspection of the images in
Fig.~\ref{fig:triplet} (including the surrounding areas) shows little
evidence for strong gravitational lensing which
could affect the observed brightness of this SN 
(we note that the host galaxy would have to be lensed as well). We cannot
conclusively rule out the possibility of strong lensing, as in the
case of PS1-10afx \citep{Quimby14}, but as discussed above the
spectrum and extended light curve of DES13S2cmm are inconsistent
with a normal SN, while the probability of such a strong lensing event
in our DES data is less than the probability of discovering a SLSN,
based on the lensing statistics given in \citet{Quimby14}, and the
lower redshift of DES13S2cmm and smaller search area of DES. Moreover, the observed $r-i$ peak colour of DES13S2cmm is consistent with the expected colours of other unlensed SNe (including SLSNe) shown in Figure 4 of \citet{Quimby14}, i.e., DES13S2cmm is located below the thick black line in this plot.
Therefore, we do not consider gravitational lensing further in this
paper, but higher-resolution imaging data should be obtained for
DES13S2cmm, after the SN event has vanished, to provide better
constraints on possible strong lensing as discussed in
\citet{Quimby14}.

%--------------------------------------------------------------------------------------------------------------------------------------------------------------------------------

\subsection{Comparison of bolometric light-curves}
\label{Bol_lc}

For comparison with DES13S2cmm, we also show in Fig.~\ref{fig:Bol_lc}
the bolometric light curves for fourteen SLSNe-I (or Ic, as discussed in \citealp{Inserra13}) in the literature: PS1-10ky. PS1-10awh \citep{Chomiuk11} and PS1-10bzj \citep{Lunnan13}; SN\,2010gx \citep{Pastorello10}; PTF\,10hgi, SN\,2011ke, PTF\,11rks, SN\,2011kf and SN\,2012il \citep{Inserra13}; SNLS\,06D4eu and SNLS\,07D2bv \citep{Howell13}; and LSQ\,12dlf, SSS\,120810 and SN\,2013dg \citep{Nicholl14}. We exclude from this figure, and our analysis, other SLSNe-I that do not possess data for at least three passbands for each epoch (defined as being taken within a 24 hour window) over a majority of their light curves.  This criterion, so defined to ensure well-constrained bolometric light-curve fits, excludes namely PTF\,09atu, PTF\,09cnd, and PTF\,09cwl \citep{Quimby11}; SCP\,06F6 \citep{Barbary09}; and SN\,2006oz \citep{Leloudas12}.

The full bolometric light curves for these SLSNe-I were calculated using the same methodology as described above for DES13S2cmm (i.e. using the published photometric data and requiring three or more passbands per epoch). As a check, our bolometric light curves for PTF\,10hgi, SN\,2011ke, PTF\,11rks, SN\,2011kf and
SN\,2012il are similar to those published in \citet{Inserra13}, with
any offsets due to differences in the integration of
the blackbody curve; we integrate over all wavelengths, while
\citet{Inserra13} integrate over the wavelength range of their observed
optical passbands (we are able to reproduce their luminosities if we
follow their procedure). Our bolometric light-curves for PS1-10ky. PS1-10awh, and PS1-10bzj are also similar to those published in the literature.

In Fig.~\ref{fig:Bol_lc}, the phase relative to peak for DES13S2cmm was defined as 
the peak of the observer-frame r-band, which occurred on MJD 56563.2, while for other 
SLSNe-I we simply used the phase relative to peak as reported by the individual 
literature studies. In the case of SNLS\,06D4eu and SNLS\,07D2bv, \citet{Howell13} reported a phase based on their calculation of the bolometric light curves, which we use herein, but note there may be greater uncertainty in the definition of the time of peak for these light curves. 

%these SLSNe are at significantly higher redshifts than the 
%others in Fig.~\ref{fig:Bol_lc}, and as such their bolometric light-curve is based 
% exclusively on data bluewards of rest-frame $g$-band.

The top panel of Fig.~\ref{fig:Bol_lc} shows that DES13S2cmm is one of the faintest SLSNe-I around peak compared to the other SLSNe-I in the literature, and possesses one of the slowest declining tails (beyond approximately $+30$ rest-frame) of the SLSNe studied herein. In the bottom panel of Fig.~\ref{fig:Bol_lc}, we compare DES13S2cmm to possible Type R SLSNe (to be consistent with the classification scheme of \citealp{Gal-Yam12}) and find the late-time light curve of DES13S2cmm is similar to both PS1-11ap \citep{2014arXiv1402.1631M} and SN2007bi \citep{Gal-Yam09}, highlighting the possible overlap between these two SLSNe types, as discussed in Section~\ref{sec:spectrum} and \citet{Inserra13}.

Fig.~\ref{fig:Bol_lc} also demonstrates that the fifteen SLSNe-I light curves (top panel) have similar luminosities at approximately $+25$ days (rest frame) past peak, close to the inflection point in the light curves when their extended 
tails begin to appear. We calculate the dispersion, as a function of relative phase, for the SLSNe-I light curves shown in Fig.~\ref{fig:Bol_lc}, linearly interpolating between large gaps in each light curve to provide an evenly sampled set of data. We find a minimum dispersion of $\sigma({\rm log_{10} L_{bol}})=0.11$ -- equivalent to $0.29$ magnitudes -- around $+30$ days past peak (rest frame), derived from the nine SLSNe in Fig.~\ref{fig:Bol_lc} constraining this phase. This dispersion in the light-curves can be decreased to $\sigma({\rm log_{10} L_{bol}})=0.083$, or $0.20$ magnitudes, at approximately $+25$ days past peak (rest frame) if we exclude the four highest redshift SLSNe-I above $z\simeq1$ (SNLS\,06D4eu, SNLS\,07D2bv, PS1-10ky. PS1-10awh). The bolometric luminosities of these SLSNe-I are more difficult to estimate because of the uncertainties associated with modelling the UV part of the spectrum by assuming a simply blackbody curve. There could also be evolution in the population with redshift.
As commented in \citet{Quimby11}, such light curve characteristics
could point towards a possible `standardisation' of SLSNe-I, as with
SNe Ia, leading to their use as high-redshift `standard candles' for
cosmological studies \citep{king14}.

Recently, \citet{inserra14} has proposed a `standardisation' of the peak magnitude of 16 SLSNe using a
stretch-luminosity correction similar to the $\Delta m_{15}$
relationship of \citet{1993ApJ...413L.105P} for SNe Ia. They find that SLSNe that are brighter at peak have slower declining light curve over the first $30$ days after peak (rest frame) and, correcting for this correlation, the
dispersion of the peak magnitude of SLSNe can be reduced to $\sigma\simeq 0.2$. This is similar in size to the dispersion between $+25$ to $+30$ days past peak seen in Fig.~\ref{fig:Bol_lc} and is only a factor of two worse than the dispersion obtained for standardised SNe Ia.

Both Fig.~\ref{fig:Bol_lc} and \citet{inserra14} suggest some standardisation of SLSNe light curves is possible but further observations and analyses are required to find the best approach. For example, the light curve of DES13S2cmm does not appear to follow the correlation found by \citet{inserra14} as it has a faint peak magnitude, relative to other SLSNe, but is also slowly declining towards $+30$ days after peak. A direct comparison to \citet{inserra14} is difficult as our study uses different sample selection criteria (only 9 of their 16 SLSNe are included here because of our requirement to have three passbands per epoch), while \citet{inserra14} estimate a common synthetic peak magnitudes for all their SLSNe using their own time-evolving spectral template. A more detailed comparison will be possible with further DES SLSNe.

%--------------------------------------------------------------------------------------------------------------------------------------------------------------------------------

\subsection{Power source of DES13S2cmm}
\label{power}

The details of the power source of SLSN-I remain unclear and are much
debated (see Section~\ref{intro}).  Here we investigate the two popular explanations in the literature for SLSNe-I: radioactive decay
of $^{56}$Ni, and energy deposition from a Magnetar.
 
%--------------------------------------------------------------------------------------------------------------

\subsubsection{Radioactive $^{56}$Ni model}
\label{power_Ni}

As can been seen in Fig.~\ref{fig:Bol_lc}, the decline rate of
DES13S2cmm is approximately that expected from the decay of $^{56}$Co
(dashed line).  This is suggestive of the SN energy source being the
production and subsequent decay of large quantities of $^{56}$Ni
produced in the explosion.  To test this theory we fit the bolometric
light curve of DES13S2cmm to a model of $^{56}$Ni, following the
prescription of \citet{Arnett82}, which is the approximation of an
homologously expanding ejecta \citep{Chatzopoulos09,Inserra13} where
the luminosity at any given time (in erg s$^{-1}$) is given by
\begin{align}	
L(t) &= M_{\textrm{Ni}} \, e^{ -(t / \tau_m)^2 }  \biggl[ \epsilon_{\textrm{Ni}} \int_{0}^{t} \frac{2u}{\tau_m^2}\,e^{(u/\tau_m)^2}\,e^{-u/\tau_{\textrm{Ni}}} du \nonumber \\
& + \epsilon_{Co} \int_{0}^{t} \frac{2u}{\tau_m^2}\,e^{(u/\tau_m)^2}\biggl(e^{-u/\tau_{\textrm{Ni}}}-e^{-u/\tau_{\textrm{Co}}}\biggr) du \biggr] \delta_{\gamma},
\end{align}		
where $u$ is the (time) integration variable. The energy production rate and decay rate for $^{56}$Ni are
$\epsilon_{\textrm{Ni}} =3.9\times10^{10}$ erg\,s$^{-1}$\,g$^{-1}$ and
$\tau_{\textrm{Ni}}=8.8$ days, while for $^{56}$Co these values are
$\epsilon_{\textrm{Co}} =6.8\times10^{9}$ erg\,s$^{-1}$\,g$^{-1}$ and
$\tau_{\textrm{Co}}=111.3$ days.  The amount of $^{56}$Ni produced in
the explosion is $M_{\textrm{Ni}}$.  Parameterizing the rise-time of the light curve,
$\tau_m$ (from Eqns 18 and 22 of \citealp{Arnett82}) is the geometric mean of the diffusion 
and expansion timescales, and is given by
\begin{equation}
\tau_m = 1.05\, \Bigl( \frac{\kappa}{\beta c} \Bigr)^{1/2} \Bigl( \frac{M^3_{\textrm{ej}}}{E} \Bigr)^{1/4}
\label{eq:tau_m}
\end{equation}
Here $E$ is the explosion energy, $M_{\textrm{ej}}$ is the total amount of ejected mass, $\kappa$ is the optical opacity (assumed to be 0.1 cm$^{2}$ g$^{-1}$ and constant throughout, as in \citet{Inserra13}), and $\beta$ is a constant with value $\approx13.7$ \citep{Arnett82}.

We use $\delta_{\gamma}$ to denote the gamma-ray deposition function: the efficiency with which gamma-rays are trapped within the SN ejecta.  For this function we follow \citet{Arnett82}, which is also used by \citet{Inserra13} and uses the deposition function defined in \citet{Colgate80},
\begin{equation}
\delta_{\gamma} = G[1+2G(1-G)(1-0.75G)],
\end{equation}
where $G\equiv\tau_{\gamma}/(\tau_{\gamma}+1.6)$, with the `optical depth' for gamma-rays approximately given by
\begin{equation}
\tau_{\gamma} \approx \Bigl( \frac{0.1}{\kappa} \Bigr) \Bigl( \frac{\tau_{m}^2}{4\tau_{\textrm{Ni}}^2} \Bigr) \Bigl( \frac{5.53 \times 10^{10}}{v_{ej}(0.1+t/\tau_{\textrm{Ni}})^2} \Bigr).
\end{equation}
Here $v_{ej} = \sqrt{10E_k/3M_{\textrm{ej}}}$ and is in units of cm s$^{-1}$.  We note that the deposition function used in \citet{Chatzopoulos09} has a functional form that is similar to that used here, but in their approximation $\tau_{\gamma}$ only deviates from $\approx1$ at much later epochs, resulting in a light-curve decay rate mirroring $^{56}$Co and largely insensitive to changes in the explosion energy or ejecta mass.

We thus have four parameters in our model: the explosion epoch
($t_0$), the energy of the explosion ($E$), the total ejected mass
(M$_{\textrm{ej}}$), and the amount of ejected mass which is $^{56}$Ni
($M_{\textrm{Ni}}$).  
We define the fraction of the ejecta mass that is in $^{56}$Ni as
$f_{\textrm{Ni}} = M_{\textrm{Ni}}/M_{\textrm{ej}}$.  
We fit our model to the data with a variety of upper limits on $f_{\textrm{Ni}}$, varying
from 0.3 to 1.0.
However, we follow \citet{Inserra13} in assuming a physically motivated upper limit on $f_{\textrm{Ni}}$ of 0.5, which they base on the prevalence of intermediate-mass elements in their spectra of lower-$z$ SLSNe, and the calculations of $^{56}$Ni production in core-collapse SNe by \citet{Umeda2008}.
We report the best-fitting parameters and goodness
of fit for these models in Table~\ref{fitparams} as well as show the range of $f_{\textrm{Ni}}$ models in Figure 6 to illustrate the scale of uncertainties in the theoretical modelling.

\begin{table}
\begin{center}
  \caption{Best-fitting parameters for a variety of $^{56}$Ni model
    fits to the bolometric light curve for DES13S2cmm.
    $f_{\textrm{Ni,max}}$ is the maximum fraction of $^{56}$Ni allowed
    in the ejecta for each fit; for DES13S2cmm the best-fitting model
    always maximizes this value.  Ejecta and $^{56}$Ni masses are
    given in units of solar masses.  There are 22 degrees of freedom
    for each fit.}
    \begin{tabular}{ c | c | c | c | c | c }	
	\hline \hline
	$f_{\textrm{Ni,max}}$ & $E(10^{51}$\,erg) & $M_{\textrm{ej}}$ & $M_{\textrm{Ni}}$ & $t_0$ (MJD) & $\chi^2$ \\
	\hline		
	0.30 & 31.88 & 14.77 & 4.43 & 56508.8 & 81.5 \\
	0.40 & 14.83 & 10.59 & 4.23 & 56510.4 & 76.0 \\
	0.50 & 8.22 & 8.21 & 4.10 & 56511.5 & 71.5 \\
	0.60 & 5.08 & 6.68 & 4.01 & 56512.4 & 67.9 \\
	0.70 & 3.38 & 5.62 & 3.93 & 56513.1 & 65.0 \\
	0.80 & 2.38 & 4.84 & 3.88 & 56513.7 & 62.7 \\
	0.90 & 1.74 & 4.25 & 3.83 & 56514.2 & 60.9 \\
	1.00 & 1.32 & 3.79 & 3.79 & 56514.6 & 59.4 \\
	\hline	
      \end{tabular}	
      \label{fitparams}
    \end{center}
  \end{table}

\begin{figure}
	\includegraphics[width=1.0\columnwidth,clip]{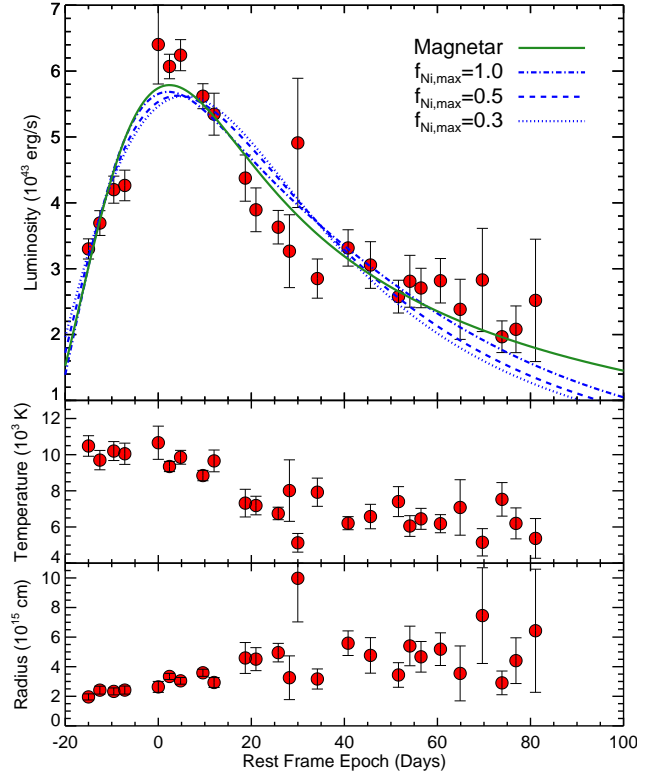}
	\caption{\textit{Upper panel:} Bolometric light curve of DES13S2cmm, with the best-fit magnetar model (green solid line) and three $^{56}$Ni models with different $f_{\rmn{Ni,max}}$ (blue lines) overplotted.  The magnetar model is a better match to the data than the best-fit $^{56}$Ni, and has a significantly better $\chi^2$ than the models with a more realistic $f_{\rmn{Ni,max}}$ (see text).  However both models have difficulty reproducing the peak luminosity, the post-peak decline, and the late-time flattening.  \textit{Middle and Bottom panels:} Temperature and radius evolution of the best-fitting blackbody to each photometric epoch which is used to construct the bolometric light curve.
          \label{fig:Models}}
\end{figure}

We find that the $^{56}$Ni model is not a good fit to the bolometric
light curve, as the best-fitting model has a $\chi^2/\rmn{dof}=2.7$
($\chi^2$ of 59 for 22 degrees of freedom), and is physically
unrealistic ($f_{\textrm{Ni}}=1$).  As can be seen in
Table~\ref{fitparams}, the model prediction for $M_\rmn{Ni}$ is
relatively robust ($3.8-4.4 M_{\odot}$), as this parameter is
primarily constrained by the peak luminosity of the light curve. 
However, to match the late time ($t>30$ days) flattening of the light
curve a large amount of ejecta mass or low explosion energy (see Eqn.~\ref{eq:tau_m}) is required to increase the diffusion time-scale, which simultaneously makes the fit to the post-peak decline poor while
predicting a much longer rise time than is seen.  We also note that these parameters are poorly constrained individually as $E$ and $M_{\textrm{ej}}$ are correlated (Equation~\ref{eq:tau_m}), and this degeneracy cannot be broken without the presence of high-quality spectral data \citep{Mazzali2013}.

We do not provide uncertainties on the best fit parameters for each $f_{\textrm{Ni}}$ model in Table~\ref{fitparams}.  This is because the formal uncertainties on parameters (such as $M_{\textrm{Ni}}$) are smaller for each fit than they are between fits with different constraints on $f_{\textrm{Ni}}$.  This is due to the simplicity of the $^{56}$Ni model; we do not consider herein Ni mixing, non-standard density profiles, nor other possible variations in this basic model that could yield different light-curve shapes.  For example, it has been noted that a two-component model that is denser in the inner component would produce a quicker flattening post-peak \citep{Maeda2003}.

%--------------------------------------------------------------------------------------------------------------

\subsubsection{Magnetar model}
\label{power_mag}

As an alternative to the radioactive decay of $^{56}$Ni, we also fit
our derived bolometric light curve for DES13S2cmm to a magnetar model,
which proposes that SLSNe are powered by the rapid spinning-down of a
neutron star with an extreme magnetic field \citep{Woosley10,Kasen10}.
In this model the input energy is defined by the initial spin
period ($P_{ms}$, in milliseconds) and magnetic field strength
($B_{14}$, in units of $10^{14}$ Gauss) of the magnetar, and the
rise-time parameter ($\tau_m$, Eqn.~\ref{eq:tau_m}) reflects the
combined effect of the explosion energy ($E$), the ejected mass
($M_{\textrm{ej}}$), and the opacity ($\kappa$).  We use the
semi-analytical model outlined in \citet{Inserra13}.  The luminosity of the 
magnetar model at time $t$ is given by (in erg s$^{-1}$) 
\begin{align}	
L(t) &= 4.9\times 10^{46}\,e^{ -(t / \tau_m)^2 }\delta \int_{0}^{t} \frac{2u}{\tau_m^2}\,e^{(u/\tau_m)^2}\,\frac{B_{14}^{2}\,P_{ms}^{-4}}{(1+u/\tau_{p})^2} du,
\end{align}
where $u$ is the (time) integration variable, $\tau_{p}$ is the spin-down timescale,
$\tau_{p}=4.7\,B_{14}^{-2}\,P_{ms}^{2}$ days, and $\delta$ is the
deposition function for the magnetar radiation; here we assume full
trapping (i.e., $\delta=1$) following \citet{Inserra13}.  We do not include any additional
contribution from $^{56}$Ni decay in this model.

In Fig.~\ref{fig:Models} we include the best-fitting magnetar model to
the bolometric light curve of DES13S2cmm.  The best-fitting parameters
of this model are a magnetic field of $1.43\times10^{14}$~Gauss, an
initial spin period of $5.28$~ms, a diffusion timescale of
$22.9$~days, and an explosion date of MJD 56511.0. This model has a
$\chi^2/\rmn{dof}=2.0$ ($\chi^2$ of 44 for 22 degrees of freedom) and is
therefore a significantly better fit to our data than the $^{56}$Ni
model.  However, it is still clear that the model does not fit
\textit{well}: it is unable to reproduce the factor of two drop in
luminosity post-peak within 30 days of explosion (although it does a
better job than $^{56}$Ni).  Intriguingly the late-time epochs seem to
fit the magnetar model well.  Our time coverage of DES13S2cmm allows
us to determine whether either model is able to reproduce detailed
features of the light curve, but discerning between the two models
would be improved by an even longer time-scale over which the two
models cannot mimic one another; DES alone will unlikely provide
longer light curve coverage as its observing season is typically not much more than 5 months a year.